\documentclass[runningheads,envcountersame]{llncs}

\usepackage{url}
\usepackage[pdftex]{hyperref}
\hypersetup{colorlinks=true,citecolor=blue,urlcolor=blue,linkcolor=blue}
\usepackage{color}

\usepackage[utf8]{inputenc}
\usepackage[T1]{fontenc}
\usepackage{xcolor}
\usepackage{booktabs}
\usepackage{amssymb}
\usepackage{amsmath}
\usepackage{cleveref}
\Crefname{figure}{Fig.}{Figs.}
\usepackage{scalerel}
\usepackage{mathtools}
\usepackage[misc]{ifsym}
\usepackage[noadjust]{cite}
\usepackage[inline]{enumitem}
\usepackage{fancyvrb}

\usepackage{fontawesome5}
\usepackage{tikz}
\usetikzlibrary{shapes,positioning,arrows}
\tikzstyle{carrow} = [-stealth]
\tikzstyle{carrowB} = [stealth-stealth]
\tikzstyle{input} = [rectangle,draw,rounded corners,fill=orange!10,
minimum width=2cm]
\tikzstyle{inference} = [circle,draw,rounded corners,fill=orange!10]
\tikzstyle{solving} = [ellipse,draw,rounded corners,fill=violet!10]
\tikzstyle{backends} = [rectangle,draw,rounded corners,fill=blue!10,
minimum width=15mm]
\tikzstyle{unkn} = [ellipse,draw,rounded corners,fill=red!45,
minimum width=2cm]
\tikzstyle{bound} = [ellipse,draw,rounded corners,fill=green!45,
minimum width=2cm]
\tikzstyle{templates} = [rectangle,draw,rounded corners,minimum width=2cm,
fill=green!10]

\newcommand{\TVar}{\mathcal{T}\Var}
\newcommand{\EC}{\mathcal{EC}}

\makeatletter
\newcommand{\superimpose}[2]{{%
    \ooalign{%
      \hfil$\m@th#1\@firstoftwo#2$\hfil\cr
      \hfil$\m@th#1\@secondoftwo#2$\hfil\cr
    }%
  }}
\makeatother

\newcommand{\smallparallel}{%
  \mathchoice%
  {\raisebox{1pt}{\scaleobj{0.6}{\parallel}}}
  {\raisebox{1pt}{\scaleobj{0.6}{\parallel}}}
  {\raisebox{0.8pt}{\scaleobj{0.45}{\parallel}}}
  {\raisebox{0.5pt}{\scaleobj{0.35}{\parallel}}}
}
\newcommand{\pto}{\mathrel{\mathpalette\superimpose{{\smallparallel\kern.1em}{\to}}}}
\newcommand{\rpto}{\mathrel{\rotatebox[origin=c]{-180}{$\mathpalette\superimpose{{\smallparallel\kern.1em}{\to}}$}}}
\newcommand{\rptoa}[1][]{\mathrel{\vphantom{\pto}^{#1}\rotatebox[origin=c]{-180}{$\mathpalette\superimpose{{\smallparallel\kern.1em}{\to}}$}}}
\newcommand{\pleadsto}{\mathrel{\mathpalette\superimpose{{\smallparallel\kern.1em}{\leadsto}}}}

\newcommand{\simto}{\stackrel{\smash{\raisebox{-.2mm}{\tiny $\sim$\kern.2em}}}{\to}}

\newcommand{\smallcirc}{%
  \mathchoice%
  {\raisebox{0.5pt}{\scaleobj{0.8}{\circ}}}
  {\raisebox{0.5pt}{\scaleobj{0.8}{\circ}}}
  {\raisebox{0.8pt}{\scaleobj{0.45}{\circ}}}
  {\raisebox{0.5pt}{\scaleobj{0.35}{\circ}}}
}
\newcommand{\mto}{\mathrel{\mathpalette\superimpose{{\smallcirc\kern.1em}{\to}}}}
\newcommand{\mleadsto}{\mathrel{\mathpalette\superimpose{{\smallcirc\kern.3em}{\leadsto}}}}

\newcommand{\m}[1]{\mathsf{#1}}
\newcommand{\mr}[1]{\mathrel{#1}}
\newcommand{\h}[1][.3]{\hspace{#1mm}}

\newcommand{\ML}[2][0]{\makebox[#1mm][l]{#2}}

\newcommand{\xR}{\mathcal{R}}
\newcommand{\xRca}{\xR_\m{ca}}
\newcommand{\xRrc}{\xR_\m{rc}}

\newcommand{\xV}{\mathcal{V}}
\newcommand{\xF}{\mathcal{F}}
\newcommand{\xT}{\mathcal{T}}

\newcommand{\xJ}{\mathcal{J}}
\newcommand{\Val}{\xV\m{al}}
\newcommand{\Var}{\xV\m{ar}}
\newcommand{\LVar}{\mathcal{L}\Var}
\newcommand{\EVar}{\mathcal{E}\Var}

\newcommand{\Dom}{\mathcal{D}\m{om}}

\newcommand{\Pos}{\mathcal{P}\m{os}}

\newcommand{\FPos}{\Pos_\xF}

\newcommand{\inter}[1]{[\![{#1}]\!]}

\newcommand{\seq}[2][n]{{#2_1},\dots,{#2_{#1}}}

\newcommand{\xFTe}{\xF_{\m{te}}}
\newcommand{\xFTh}{\xF_{\m{th}}}
\newcommand{\SET}[1]{\{\h#1\h\}}

\newcommand{\CO}[1]{[\h#1\h]}

\newcommand{\R}{\rightarrow}
\newcommand{\Ra}[1][]{\R^{#1}}
\newcommand{\Rb}[1][]{\R_{#1}}
\newcommand{\Rab}[2][]{\R_{#1}^{#2}}
\newcommand{\RbR}{\Rb[\xR]}

\newcommand{\xRa}[1][]{\xrightarrow{#1}}

\renewcommand{\L}{\leftarrow}
\newcommand{\La}[1][]{\mr{\vphantom{\R}^{#1}{\L}}}
\newcommand{\Lb}[1][]{\mr{\vphantom{\R}_{#1}{\L}}}

\newcommand{\J}{\downarrow}
\newcommand{\Jb}[1][]{\J_{#1}}

\newcommand{\C}{\leftrightarrow}

\newcommand{\CCP}{\m{CCP}}
\newcommand{\CPCP}{\m{CPCP}}
\newcommand{\overlap}[3][p]{\langle #2, #1, #3 \rangle}
\newcommand{\crr}[3]{#1 \R #2~\CO{#3}}
\newcommand{\CRR}{\crr{\ell}{r}{\varphi}}

\newcommand{\MC}[2][0]{\makebox[#1mm]{#2}}
\newcommand{\RB}[2][1]{\raisebox{#1mm}{#2}}
\newcommand{\Rs}{\stackrel{\smash{\RB[-.5]{\tiny $\sim$~}}}{\R}}
\newcommand{\sRb}[1]{\Rs_{#1}}
\newcommand{\sRbR}{\sRb{\xR}}
\newcommand{\sRab}[2][]{\Rs_{#1}^{#2}}

\newcommand{\MS}[1][]{\xRa[\smash{%
\RB[-2]{\MC[2.5]{$\stackrel{#1}{\circ}$}}}]}

\newcommand{\sMS}{\stackrel{\smash{\RB[-1.2]{\tiny $\sim$\:}}}{\MS}}
\newcommand{\sMSab}[2][]{\sMS_{#1}^{#2}}
\newcommand{\spto}{\stackrel{\smash{\RB[-1.0]{\tiny $\sim$\:}}}{\pto}}
\newcommand{\srpto}{\stackrel{\smash{\kern.2em\RB[-0.7]{\tiny
$\sim$\:}}}{\rpto}}
\newcommand{\sptoab}[2][]{\mr{\smash{\spto}_{#1}^{#2}}}
\newcommand{\crest}{\textsf{crest}}
\newcommand{\crestcade}{\textsf{prototype}}
\newcommand{\Ctrl}{\textsf{Ctrl}}
\newcommand{\Cora}{\textsf{Cora}}
\newcommand{\CSI}{\textsf{CSI}}
\newcommand{\RMT}{\textsf{RMT}}
\newcommand{\CRaris}{\textsf{CRaris}}
\newcommand{\tct}{\textsf{TcT}}
\newcommand\tttt{%
 \textsf{T\kern-0.15em\raisebox{-0.55ex}T\kern-0.15emT\kern-0.15em\raisebox{-0.55ex}2}%
}

\usepackage{pifont}
\newcommand{\Y}{\ding{51}}
\newcommand{\N}{\ding{55}}
\newcommand{\M}{\textbf{?}}
\newcommand{\T}{\textbf{\dag}}

\begin{document}

\title{Automated Analysis of Logically Constrained Rewrite Systems
using \crest\thanks{This research is funded by the Austrian
Science Fund (FWF) project I5943.}}
\titlerunning{Automated Analysis of LCTRSs using \crest}

\author{Jonas Sch\"opf%
\textsuperscript{(\href{mailto:jonas.schoepf@uibk.ac.at}{\Letter})}%
\orcidID{0000-0001-5908-8519} \and
Aart Middeldorp\orcidID{0000-0001-7366-8464}}
\authorrunning{J. Sch\"opf, A. Middeldorp}

\institute{Department of Computer Science, University of Innsbruck,
Innsbruck, Austria
\email{\{jonas.schoepf,aart.middeldorp\}@uibk.ac.at}}

\maketitle

\begin{abstract}
We present \crest, a tool for automatically proving
(non-) confluence and termination of logically constrained rewrite
systems. We compare \crest{} to other tools for logically constrained
rewriting. Extensive experiments demonstrate the promise of \crest.
\end{abstract}

\section{Introduction}
\label{sec:introduction}

Term rewriting is a simple Turing-complete model of computation.
Properties like confluence and termination are of key interest and
numerous powerful tools have been developed for their analysis.
Logically constrained term rewrite systems (LCTRSs for short) constitute a
natural extension of term rewrite systems (TRSs) in which rules are
equipped with
logical constraints that are handled by SMT solvers, thereby avoiding
the cumbersome encoding of operations on e.g.\ integers and bit vectors
in term rewriting. LCTRSs, introduced by Kop and Nishida in
2013~\cite{KN13}, are useful for program
analysis~\cite{CL18,FKN17,KN23,WM21}. They also developed
\Ctrl~\cite{KN14,KN15}, a tool for LCTRSs
specializing in termination analysis and equivalence testing. Later,
techniques for completion~\cite{WM18} and non-termination~\cite{NW18}
analysis were added.

In this paper we describe \crest, the Constrained REwriting Software Tool.
The tool \crest{} was first announced in \cite{SM23} with support for a
small number of confluence techniques. The new version described here
includes numerous extensions:
\begin{itemize}
\item
more advanced confluence techniques (introduced in \cite{SMM24}),
\smallskip
\item
automated non-confluence and termination analysis,
\smallskip
\item
support for fixed-sized bit vectors,
\smallskip
\item
transformation techniques based on splitting critical pairs and merging
constrained rewrite rules, to further boost the confluence proving
power.
\end{itemize}
Extensive experiments show the strength of \crest.
The tool is open-source and available from
\url{http://cl-informatik.uibk.ac.at/software/crest}.

The remainder of the paper is organized as follows. In the next section
we recall important definitions pertaining to LCTRSs. \Cref{sec:techniques}
summarizes the main confluence and termination techniques implemented
in \crest. Automation details are presented in \Cref{sec:automation}. The
new transformation techniques are described in \Cref{sec:transformation}.
In \Cref{sec:evaluation} we present our experiments, before concluding in
\Cref{sec:conclusion} with suggestions for future extensions.
We conclude this introductory section with mentioning other tools
for LCTRSs.

\paragraph{Related Tools.}

We already mentioned \Ctrl%
\footnote{\url{http://cl-informatik.uibk.ac.at/software/ctrl/}}
which until 2023 was the only tool capable
of analyzing confluence and termination of LCTRSs. It supports
termination analysis~\cite{K16}, completion techniques~\cite{WM18},
rewriting induction for equivalence testing of LCTRSs~\cite{FKN17},
and basic confluence analysis~\cite{KN15}.
Unfortunately, it is neither actively maintained nor very well documented,
which is one reason why the development of \crest{} was started.
Moreover, a branch%
\footnote{\url{https://github.com/bytekid/tct-lctrs}}
of the automated resource analysis tool \tct~\cite{AMS16}
performs complexity analysis on LCTRSs based on~\cite{WM21}.
\RMT{} by Ciob\^{a}c\u{a} et al.~\cite{CL18,CLB23} is a newer
tool for program analysis based on a variation of LCTRSs.

In the 2024 edition of the Confluence Competition%
\footnote{\url{https://ari-cops.uibk.ac.at/CoCo/2024/competition/LCTRS/}}
the tool
\CRaris,\footnote{\url{https://www.trs.css.i.nagoya-u.ac.jp/craris/}}
developed by Nishida and Kojima, made its appearance.
The tool implements weak orthogonality~\cite{KN13} and the
Knuth--Bendix criterion for
terminating LCTRSs~\cite{SM23}. For termination,
it implements the dependency pair framework~\cite{K16} and the
singleton self-looping removal processor~\cite{MNKS23-arxiv} for LCTRSs
with bit vectors.

Also in 2024 Guo et al.~\cite{GK24,GHKV24} announced \Cora{}, a new
open-source tool for termination analysis of logically constrained
\emph{simply-typed} term rewrite systems, which serve as a high-order
generalization of LCTRSs. It employs static dependency pairs~\cite{KS07}
with several base methods, including a variant of the higher-order
recursive path order~\cite{JR99}.

\section{Logically Constrained Term Rewriting}
\label{sec:background}

Familiarity with the basic notions of term
rewriting~\cite{BN98} is assumed.
We assume a many-sorted signature
$\xF = \xFTe \cup \xFTh$ consisting of term and theory symbols
together with a countably infinite set of variables $\xV$. For
every sort $\iota$ in $\xFTh$ we have a
non-empty set $\Val_\iota \subseteq \xFTh$ of value
symbols, such that all $c \in \Val_\iota$ are constants of sort $\iota$.
We demand $\xFTe \cap \xFTh \subseteq \Val$ where
$\Val = \bigcup_\iota \Val_\iota$. 
The set of terms constructed from function symbols in $\xF$ and
variables in $\xV$ is by $\xT(\xF,\xV)$.
A term in $\xT(\xFTh,\xV)$ is called a \emph{logical} term.
Ground logical terms are mapped to values by an
interpretation $\xJ$:
$\inter{f(\seq t)} = f_\xJ(\inter{t_1},\dots,\inter{t_n})$.
Logical terms of sort $\m{bool}$ are called \emph{constraints}.
A constraint $\varphi$ is \emph{valid} if
$\inter{\varphi\gamma} = \top$ for all substitutions $\gamma$ such that
$\gamma(x) \in \Val$ for all $x \in \Var(\varphi)$.
Positions are sequences of positive integers to indicate
subterms. The root of a term is denoted by the empty string $\epsilon$.
For a term $s$, its subterm at position $p$ is given by $s|_p$.
The set of positions in $s \in \xT(\xF,\xV)$ is denoted by $\Pos(s)$
whereas $\FPos(s)$ is
restricted to positions with function symbols in $s$. We write
$\Var(s)$ for the set of variables in $s$.
A \emph{constrained rewrite rule} is a triple $\rho\colon \CRR$ where
$\ell, r \in \xT(\xF,\xV)$ are terms of the same sort such that
$\m{root}(\ell) \in \xFTe \setminus \xFTh$ and $\varphi$ is a
constraint.
We denote the set $\Var(\varphi) \cup (\Var(r) \setminus \Var(\ell))$ of
\emph{logical} variables in $\rho$ by $\LVar(\rho)$.
We write $\EVar(\rho)$ for the set
$\Var(r) \setminus (\Var(\ell) \cup \Var(\varphi))$ of \emph{extra}
variables.
A set of constrained rewrite rules is called an LCTRS.
A substitution $\sigma$
\emph{respects} a rule
$\rho\colon \CRR$, denoted by $\sigma \vDash \rho$, if
$\Dom(\sigma) \subseteq \Var(\rho)$,
$\sigma(x) \in \Val$ for all $x \in \LVar(\rho)$, and
$\varphi\sigma$ is
valid. Moreover, a constraint $\varphi$ is respected by $\sigma$,
denoted by $\sigma \vDash \varphi$, if $\sigma(x) \in \Val$ for all
$x \in \Var(\varphi)$ and $\varphi\sigma$ is valid. We call
$f(\seq{x}) \R y~\CO{y = f(\seq{x})}$ with a fresh variable $y$
and $f \in \xFTh \setminus \Val$ a \emph{calculation rule}. The set of
all calculation rules induced by the signature $\xFTh$ of an LCTRS
$\xR$ is denoted by $\xRca$ and we abbreviate $\xR \cup \xRca$ to
$\xRrc$. A rewrite step $s \RbR t$ satisfies $s|_p = \ell\sigma$ and
$t = s[r\sigma]_p$ for some position $p$, constrained rewrite rule
$\rho\colon \CRR$ in $\xRrc$,
and substitution $\sigma$ such that $\sigma \vDash \rho$.

A \emph{constrained term} is a pair $s~\CO{\varphi}$ consisting
of a term $s$ and a constraint $\varphi$. Two constrained terms
$s~\CO{\varphi}$ and $t~\CO{\psi}$ are \emph{equivalent},
denoted by $s~\CO{\varphi} \sim t~\CO{\psi}$, if for every substitution
$\gamma \vDash \varphi$ with $\Dom(\gamma) = \Var(\varphi)$
there is some substitution $\delta \vDash \psi$ with
$\Dom(\delta) = \Var(\psi)$ such that $s\gamma = t\delta$, and vice versa.
Let $s~\CO{\varphi}$ be a constrained term.
If $s|_p = \ell\sigma$ for some constrained rewrite rule
$\rho\colon \crr{\ell}{r}{\psi} \in \xRrc$, position $p$, and
substitution $\sigma$ such that $\sigma(x) \in \Val \cup \Var(\varphi)$
for all $x \in \LVar(\rho)$, $\varphi$ is satisfiable and
$\varphi \Rightarrow \psi\sigma$ is valid then
$s~\CO{\varphi} \RbR s[r\sigma]_p~\CO{\varphi}$.
The rewrite relation $\sRbR$ on constrained terms is defined as
$\sim \cdot \RbR \cdot \sim$ and
$s~\CO{\varphi} \sRab[\xR]{p} t~\CO{\psi}$
indicates that the rewrite step in $\sRbR$ takes place at position
$p$ in $s$. Similarly, we write
$s~\CO{\varphi} \sRb{\geqslant p} t~\CO{\psi}$
if the position in the rewrite step is below position $p$.
We illustrate some of these concepts by means of a simple example
which models the computation of the maximum of two integers.

\begin{example}
\label{exa:prelim-rewriting}
Consider the LCTRS $\xR$ over the theory \textsf{Ints} with the rules
\begin{align*}
\alpha\colon \m{max}(x,y) &\R x~\CO{x \geqslant y} &
\beta\colon \m{max}(x,y) &\R y ~\CO{y \geqslant x}
\end{align*}
Here $x$ and $y$ are logical variables in both rules. There are no
extra variables. The symbol $\m{max}$ is the only term symbol. The theory
symbols depend on the definition of \textsf{Ints}.
As the goal is automation this usually consists of non-linear integer
arithmetic as specified in the respective SMT-LIB theory.%
\footnote{\url{https://smtlib.cs.uiowa.edu/Theories/Ints.smt2}}

By applying the calculation rule $x_1 + x_2 \R y~\CO{y = x_1 + x_2}$
with substitution
$\SET{x_1 \mapsto \m{3}, x_2 \mapsto \m{2}, y \mapsto \m{5}}$
followed by rule $\alpha$ we obtain
\begin{align*}
\m{max}(\m{3} + \m{2},\m{3}) &\R \m{max}(\m{5},\m{3}) \R \m{5}
\intertext{An example of constrained rewriting is given by}
\m{max}(\m{3},\m{3} + x)~\CO{x \geqslant \m{0}}
&\Rs \m{max}(\m{3},z)~\CO{x \geqslant \m{0} \land z = \m{3} + x} \\
&\Rs z~\CO{x \geqslant \m{0} \land z = \m{3} + x}
\end{align*}
\end{example}

One-step rewriting, i.e., rewriting a term using a single rule, was
introduced above.
The sufficient criteria for confluence, highlighted
in the next section, heavily rely on the notation of parallel ($\pto$) and
multi-step ($\mto$) rewriting following \cite[Definition~3]{SMM24} and
\cite[Definition~8]{SM23}. The former is capable of applying
several rules at parallel positions in a step while the latter
additionally allows recursive steps within the used matching
substitutions of rules. 
A rewrite sequence consists of consecutive rewrite steps,
independent of which kind.
The reflexive and transitive closure of $\R$ is denoted by $\Ra[*]$.
Moreover, for arbitrary terms $s$ and $t$ we write $s \C t$
if $s \mr{({\L} \cup {\R})} t$ and $s \J t$ if there exists a
term $u$ such that $s \Ra[*] u \La[*] t$.

\section{Confluence and Termination}
\label{sec:techniques}

Termination and confluence are well-known properties in static program
analysis. Both properties are in general undecidable. With respect to
(logically constrained) term rewriting, a program is terminating whenever
it does not admit an infinite rewrite sequence.
Confluence states that $s \J t$ whenever $t \La[*] s \Ra[*] u$,
for all terms $s$, $t$ and $u$.
Naively checking the properties is obviously not feasible.
In (logically constrained) term rewriting
there exist sufficient criteria
that guarantee that these properties are satisfied for
a given program.
In the following we highlight key components of confluence and termination
analysis for
logically constrained rewrite systems.

The confluence methods implemented in \crest{} are based on
(parallel) critical pairs. These are defined as follows.
Given a constrained rewrite rule $\rho$, we write
$\EC_\rho$ for $\bigwedge \SET{x = x \mid x \in \EVar(\rho)}$.
An \emph{overlap} of an LCTRS $\xR$ is a triple $\overlap{\rho_1}{\rho_2}$
with rules $\rho_1\colon \crr{\ell_1}{r_1}{\varphi_1}$ and
$\rho_2\colon \crr{\ell_2}{r_2}{\varphi_2}$, satisfying the following
conditions: (1) $\rho_1$ and $\rho_2$ are variable-disjoint variants of
rewrite rules in $\xRrc$, (2) $p \in \FPos(\ell_2)$, (3) $\ell_1$ and
$\ell_2|_p$ unify with mgu $\sigma$ such that
$\sigma(x) \in \Val \cup \xV$ for all
$x \in \LVar(\rho_1) \cup \LVar(\rho_2)$, (4)
$\varphi_1\sigma \land \varphi_2\sigma$ is satisfiable, and (5) if
$p = \epsilon$ then $\rho_1$ and $\rho_2$ are not variants, or
$\Var(r_1) \nsubseteq \Var(\ell_1)$. In this case we call
$\ell_2\sigma[r_1\sigma]_p \approx r_2\sigma~
\CO{\varphi_1\sigma \land \varphi_2\sigma \land \psi\sigma}$
a \emph{constrained critical pair} (CCP) obtained from the overlap
$\overlap{\rho_1}{\rho_2}$. Here
$\psi = \EC_{\rho_1} \land \EC_{\rho_2}$.
The peak
\begin{gather*}
\ell_2\sigma[r_1\sigma]_p~\CO{\Phi} \L \ell_2\sigma~\CO{\Phi}
\R^\epsilon r_2\sigma~\CO{\Phi}
\end{gather*}
with $\Phi = (\varphi_1 \land \varphi_2 \land \psi)\sigma$,
from which the constrained critical pair originates, is called a
\emph{constrained critical peak}. The set of all constrained critical
pairs of $\xR$ is denoted by $\CCP(\xR)$.
A constrained equation $s \approx t~\CO{\varphi}$ is \emph{trivial} if
$s\sigma = t\sigma$ for every substitution $\sigma$ with $\sigma \vDash
\varphi$. The trivial equations from $\EC_{\rho}$ are used in order to
prevent loosing the information which (extra) variables are logical
variables in the underlying rules of a CCP.

\begin{example}
\label{exa:prelim-confluence}
Let us extend the LCTRS $\xR$ from \Cref{exa:prelim-rewriting}
with an additional rule modeling the commutativity of $\m{max}$:
\begin{align*}
\m{max}(x,y) &\R x~\CO{x \geqslant y} &
\m{max}(x,y) &\R y~\CO{y \geqslant x} &
\m{max}(x,y) &\R \m{max}(y,x)
\end{align*}
There are six constrained critical pairs, including the following two:
\begin{align*}
x &\approx y~\CO{x \geqslant y \land y \geqslant x} &
x &\approx \max(y,x)~\CO{x \geqslant y}
\end{align*}
The left one is trivial, the one on the right becomes trivial after one
rewrite step:
$x \approx \max(y,x)~\CO{x \geqslant y} \R x \approx x~\CO{x \geqslant y}$.
The remaining four pairs can be rewritten similarly.
\end{example}

Restricting the way in which constrained critical peaks are rewritten
into trivial ones, yields different sufficient conditions for confluence
of (left-)linear LCTRSs. We state the conditions below but refer
to~\cite{KN13,SM23,SMM24} for precise definitions:
\begin{enumerate}[label=(C\arabic*)]
\item
\label{cr-wo}
(weak) orthogonality (\cite[Theorem~4]{KN13}),
\smallskip
\item
\label{cr-kb}
joinable critical pairs for terminating LCTRSs (\cite[Corollary 4]{SM23}).
\smallskip
\item
\label{cr-sc}
strong closedness for linear LCTRSs (\cite[Theorem~2]{SM23}),
\smallskip
\item
\label{cr-apc}
(almost) parallel closedness for left-linear LCTRSs
(\cite[Theorem~4]{SM23}),
\smallskip
\item
\label{cr-adc}
(almost) development closedness for left-linear LCTRSs
(\cite[Corollary~1]{SMM24}).
\end{enumerate}

The final confluence criterion implemented in \crest{} is based on
parallel critical pairs.
Let $\xR$ be an \textup{LCTRS}, $\rho\colon \ell \R r~\CO{\varphi}$ a
rule in $\xRrc$, and $P \subseteq \FPos(\ell)$ a non-empty set of parallel
positions. For every $p \in P$ let
$\rho_p\colon \ell_p \R r_p~\CO{\varphi_p}$ be a variant of a rule in
$\xRrc$. Let $\psi = \EC_{\rho} \land
\bigwedge_{p \in P} \EC_{\rho_p}$ and $\Phi = \varphi\sigma \land
\psi\sigma \land \bigwedge_{p \in P} \varphi_p\sigma$. The peak
$\ell\sigma[r_p\sigma]_{p \in P}~\CO{\Phi} \rpto \ell\sigma~\CO{\Phi}
\Rab[\xR]{\epsilon} r\sigma~\CO{\Phi}$ forms a
\emph{constrained parallel critical pair}
$\ell\sigma[r_p\sigma]_{p \in P} \approx r\sigma~\CO{\Phi}$ if the
following conditions are satisfied:
\begin{enumerate}
\item
$\Var(\rho_1) \cap \Var(\rho_2) = \varnothing$ for different rules
$\rho_1$ and $\rho_2$ in $\SET{\rho} \cup \SET{\rho_p \mid p \in P}$,
\item
$\sigma$ is an mgu of $\SET{\ell_p = \ell|_p \mid p \in P}$
such that $\sigma(x) \in \Val \cup \xV$ for all
$x \in \LVar(\rho) \cup \bigcup_{p \in P} \LVar(\rho_p)$,
\item
$\varphi\sigma \land \bigwedge_{p \in P} \varphi_p\sigma$ is satisfiable,
and
\item
if $P = \SET{\epsilon}$ then $\rho_\epsilon$ is not a variant of $\rho$
or $\Var(r) \nsubseteq \Var(\ell)$.
\end{enumerate}
A constrained peak forming a constrained parallel critical pair is
called a \emph{constrained parallel critical peak}. The set of all
constrained parallel critical pairs of $\xR$ is denoted by $\CPCP(\xR)$.
The following sufficient condition for confluence is reported in
(\cite[Corollary~2]{SMM24}):
\begin{enumerate}[label=(C\arabic*)]
\addtocounter{enumi}{5}
\item
\label{cr-pcpcp}
parallel closedness of parallel critical pairs for left-linear LCTRSs.
\end{enumerate}

Conditions \ref{cr-adc} and \ref{cr-pcpcp} do not subsume each other. Both
generalize conditions \ref{cr-wo} -- \ref{cr-apc}.
All these confluence criteria try to find a specific closing rewrite
sequence starting from a constrained (parallel) critical
pair---which is seen as a constrained equation---to
a trivial constrained equation. For example, parallel
closedness in~\ref{cr-apc} involves showing that each constrained critical
pair $s \approx t~\CO{\varphi}$ can be rewritten into a trivial constrained
equation using a single parallel step
$s \approx t~\CO{\varphi} \sptoab[\geqslant 1]{}
s' \approx t~\CO{\varphi}$.
Note that only the left part ($s$) is rewritten here and
$s' \approx t~\CO{\varphi}$ is a trivial constrained equation.
For (finite) terminating TRSs, confluence is decided by rewriting
critical pairs to normal form~\cite{KB70}. For terminating LCTRSs
confluence---even for a decidable theory---is undecidable~\cite{SMM24},
but rewriting constrained critical pairs to normal forms is still of
value. This is used in~\ref{cr-kb} above.
We need however to adapt the notion of normal form for constrained
terms.

\begin{example}
\label{exa:example 3}
The LCTRS $\xR$ over the theory $\m{Ints}$ with rewrite rules
\begin{align*}
\m{f}(x) &\R \m{g}(x)~\CO{x \geqslant \m{1}} &
\m{g}(\m{1}) &\R \m{a} &
\m{h}(x) &\R \m{a}~\CO{x \leqslant \m{1}} \\
\m{f}(x) &\R \m{h}(x)~\CO{x \leqslant \m{2}} &
\m{g}(x) &\R \m{b}~\CO{x \geqslant \m{2}} &
\m{h}(x) &\R \m{b}~\CO{x > \m{1}} \\
&& \m{g}(x) &\R \m{c}~\CO{x < \m{1}}
\end{align*}
admits one (modulo symmetry) constrained critical pair:
\begin{gather*}
\m{g}(x) \approx \m{h}(x)~\CO{x \geqslant \m{1} \land x \leqslant \m{2}}
\end{gather*}
None of the rules above are applicable, so this non-trivial constrained
critical pair is in normal form with respect to $\RbR$, but it would be
wrong to conclude that $\xR$ is not confluent; all substitutions $\sigma$
that satisfy the constraint $x \geqslant \m{1} \land x \leqslant \m{2}$
allow us to rewrite $(\m{g}(x) \approx \m{h}(x))\sigma$ to the trivial
equations $\m{a} \approx \m{a}$ or $\m{b} \approx \m{b}$.
\end{example}

\begin{definition}
\label{def:cnf}
Given an LCTRS $\xR$, a constrained term $s~\CO{\varphi}$ is in
\emph{normal form} if and only if for all substitutions $\sigma$ with
$\sigma \vDash \varphi$ we have $s\sigma \RbR t$ for no term $t$.
\end{definition}

Note that the constrained critical pair in \Cref{exa:example 3}
is not in normal form according to this definition. We present a simple
sufficient condition for non-confluence. The easy proof is given in
the appendix.

\begin{lemma}
\label{lem:non-confluence}
An \textup{LCTRS} is \emph{non-confluent} if there exists a
constrained critical pair
that rewrites to a non-trivial constrained equation in normal form.
\end{lemma}

We will resume the analysis of \Cref{exa:example 3} in
\Cref{sec:transformation}.
Termination plays an important role in the analysis of LCTRSs.
\crest{} implements the following methods reported in the papers by
Kop and Nishida~\cite{KN13,K16}:
\begin{enumerate}[label=(T\arabic*)]
\item
\label{sn-dpg}
dependency graph (\cite[Theorems~4 \& 5]{K16}),
\smallskip
\item
\label{sn-rpo}
recursive path order (\cite[Theorem~5]{KN13}),
\smallskip
\item
\label{sn-vc}
value criterion (\cite[Theorem~10]{K16}),
\smallskip
\item
\label{sn-rpair}
reduction pairs (\cite[Theorem~12]{K16}).
\end{enumerate}
Method \ref{sn-dpg} computes the strongly connected components in
the dependency graph, and transforms the input LCTRS into
so-called DP problems, which can be analyzed independently. It lies at
the heart of the dependency pair framework~\cite{GTS05} implemented in most termination
tools for TRSs.
Methods \ref{sn-rpo} and \ref{sn-rpair} are LCTRS variants of
well-known methods for TRSs~\cite{D82,AG00}.
Two further methods implemented in \Ctrl{} are ported to \crest:
\begin{enumerate}[label=(T\arabic*)]
\addtocounter{enumi}{4}
\item
\label{sn-sc}
subterm criterion
\smallskip
\item
\label{sn-vclin}
special value criterion
\end{enumerate}
While \ref{sn-sc} is a well-known termination method for DP problems
originating from TRSs~\cite{HM04}, \ref{sn-vc} and \ref{sn-vclin} are
specific to LCTRSs. Method \ref{sn-sc} operates on the syntactic
structure of dependency pairs and ignores the constraints.
In method \ref{sn-vc} dependency pair symbols are also
projected to a direct argument
but then a strict decrease with respect to the constraint is required.
For example, the rule $\m{f}(x) \R \m{f}(x - 1)~\CO{x > 0}$ cannot be
handled by \ref{sn-sc}, 
but as $x > 0$ implies the strict decrease $x \succ x - 1$
for a suitable well-founded relation $\succ$, \ref{sn-vc} applies.
Method \ref{sn-vclin} is an extension of \ref{sn-vc} in which
linear combinations of arguments are considered.
Methods \ref{sn-vc} and \ref{sn-vclin} are adapted to the higher-order
LCTRS setting in \cite[Sections~4.2 \& 4.3]{GHKV24}.

\section{Automation}
\label{sec:automation}

Our tool \crest{} is written in Haskell and the current version consists
of roughly 12000 lines of code. Core modules like SMT solving use a fork
of the \texttt{simple-smt} package%
\footnote{\url{https://hackage.haskell.org/package/simple-smt}}
and the rewriting modules are
inspired by the \texttt{term-rewriting}
package.%
\footnote{\url{https://hackage.haskell.org/package/term-rewriting}}
In the following we provide some details of the key components.

\paragraph{Input Format.}
\crest{} operates on LCTRSs in the new ARI
format\footnote{\label{foot:format}%
\url{https://project-coco.uibk.ac.at/ARI/lctrs.php}}~\cite{AHKKMMNSSSTY23}
adopted by the Confluence Competition (CoCo) and also
partly by the Termination Competition.%
\footnote{\url{https://termination-portal.org/wiki/Termination_Portal}}
Problems in the ARI database are given a unique number which we will
use throughout this paper to address specific LCTRSs. An example problem
is given in \Cref{fig:ari-adc}.
The ARI database format requires sort annotations for
variables appearing as an argument of a polymorphic predicate. If this
sort can be inferred at a different position then this can be ignored for
\crest. For example, consider the rule
$\m{f}(x = y,x) \R z~\CO{z = x + \m{1}}$ with
$\m{f}\colon \m{Bool} \to \m{Int} \to \m{Int}$,
${+}\colon \m{Int} \to \m{Int} \to \m{Int}$ and
${=}\colon A \to A \to \m{Bool}$
with a polymorphic sort $A$. In the ARI database all variables need
concrete sort annotation. For \crest{} no sort annotation is
necessary as all the sorts of variables can be inferred from the sort of
$\m{f}$.

\begin{figure}[tb]
\centering
\begin{Verbatim}[fontsize=\small]
(format LCTRS :smtlib 2.6)
(theory Ints)
(fun f (-> Int Int Int))
(fun g (-> Int Int Int))
(fun c (-> Int Int Int))
(fun h (-> Int Int))
(rule (f x y) (h (g y (* 2 2))) :guard (and (<= x y) (= y 2)))
(rule (f x y) (c 4 x) :guard (<= y x))
(rule (g x y) (g y x))
(rule (c x y) (g 4 2) :guard (not (= x y)))
(rule (h x) x)
\end{Verbatim}
\vspace{-2ex}
\caption{ARI file \href{https://ari-cops.uibk.ac.at/ARI/%
?m=problems\&d=ARI\&q=1528}{1528} (without sort annotations and meta
information).}
\label{fig:ari-adc}
\end{figure}

Theory symbols are those that are defined in a specific SMT-LIB theory,
however, for fixed-sized bit vectors \crest{} additionally supports
function symbols defined in the SMT-LIB logic
\texttt{QF\textunderscore{}BV}.%
\footnote{\url{https://smt-lib.org/logics-all.shtml\#QF_BV}}
In addition to LCTRSs also plain TRSs and many-sorted TRSs are
supported.

\paragraph{Pre-Processing.}
After parsing its input and assigning already known sorts to function
symbols and variables we apply a basic type inference algorithm. Some
function symbols in the core theory, which provides basic boolean
functions, like ``$=$'' have a polymorphic sort. Therefore we need to
infer unknown sorts in order to obtain a fully sorted LCTRS. This is
required as sort information must be present
for the declaration of variables in the SMT solver. During the parsing
phase \crest{}
parses the respective theory from an internal representation of the
SMT-LIB specification.
Currently the theory of integers, reals, fixed-sized bit vectors and a
combination of integers and reals are supported.
Subsequently \crest{} preprocesses the LCTRS by moving values
in the left-hand sides of the rewrite rules into the constraints (by
applying the transformation described in~\cite[Definition~13]{SM23}).
Afterwards it merges as many rules as possible following
\Cref{def:merge-rules} in \Cref{sec:transformation}.

\paragraph{Rewriting.}
One of the key components is the rewriting module which provides
functionality to perform rewriting on constrained terms.
This module computes
rewrite sequences of arbitrary length, using single steps, parallel
rewrite steps~\cite[Definition~7]{SMM24} and
multisteps~\cite[Definition~5]{SMM24}.
Calculation steps are modeled in an obvious way; whenever
we have a term $s[f(\seq{s})]~\CO{\varphi}$ with
$\seq{s} \in \Val \cup \Var(\varphi)$ and $f \in \xFTh$, then we produce
$s[x]~\CO{\varphi \land x = f(\seq{s})}$ for a fresh
variable $x$. In some cases single rule steps need more care because of
the lack of equivalence steps in rewrite sequences.
For rules with variables that do not occur in the left-hand side, the
matching substitution of the left-hand side does not provide an
instantiation. However, those variables are logical and need to be
instantiated with values. This is achieved by adding the constraint of
the rule and its extra variables to
the resulting constrained term after it has been confirmed that for those
variables an instantiation exists. We illustrate this in the
following example.

\begin{example}
Consider the constrained rule
$\rho\colon \m{f}(x) \R y~\CO{x \geqslant 0 \land x > y}$, the constrained
term $\m{f}(z)~\CO{z = \m{2}}$
and the matching substitution $\SET{x \mapsto z}$ between the left-hand
side of $\rho$ and $\m{f}(z)$.
The variable $y$ is not part of the matching substitution and thus
\crest{} rewrites $\m{f}(z)~\CO{z = \m{2}}$ to
$y~\CO{z = \m{2} \land z \geqslant 0 \land z > y}$. Using the constrained
rule $\rho'\colon \m{f}(x) \R y~\CO{x \geqslant 0}$ from the same
constrained term would give
$y~\CO{z = \m{2} \land z \geqslant 0 \land y = y}$.
\end{example}

\paragraph{SMT.}
SMT solving is a key component in the analysis of LCTRSs and SMT solvers
are heavily used during the analysis. In order for SMT solving to not
form a bottleneck some care is needed. Again, each different analysis
method is equipped with its own SMT solver instance started at the
beginning of the analysis. Afterwards such an instance runs until the
method has finished. In between, it waits for SMT queries, hence we avoid
several restarts of this instance. Constraints are modeled as regular
terms of sort boolean and can be checked for satisfiability and validity.
Each of those checks runs in its own context (using push and pop commands)
in order to avoid any interference with previous queries.
Currently \crest{} utilizes \textsf{Z3}~\cite{dMB08} as the default
SMT solver, as it turned out to be the most reliable during development.
Nevertheless, \crest{} provides the (experimental) possibility to use
\textsf{Yices}~\cite{D14} and \textsf{CVC5}~\cite{BBBKLMMMNNOPRSTZ22}.

\paragraph{Confluence.}
The computation of constrained critical pairs follows the definition and
constrained parallel critical pairs are computed in a bottom up fashion
by collecting all possible combinations of parallel steps. Then the
various methods to conclude confluence are
applied on those pairs. If a method fails on
a constrained critical pair then, using \Cref{def:split-ccp}, the
constrained critical pair is split.
The logical constraint used in splitting is taken from a matching rule.
The various methods run concurrently in order to prevent
starvation of methods because of pending SMT solver queries.
The first method which succeeds returns
the result and all others, including their SMT solver instances, are
terminated. We adopt heuristics to bound the number of rewrite steps
in the closing sequences.
The method that posed the biggest challenge
to automation is the 2-parallel closedness~\cite[Definition~11]{SMM24}
needed for \ref{cr-pcpcp} as we cannot
simply use an arbitrary parallel step starting from the right-hand side
but need to synthesize a parallel step over a set of parallel positions
that adheres to the variable condition present in the definition.

\paragraph{Termination.}
The choices in the parameters of the subterm criterion \ref{sn-sc} and the
recursive path order \ref{sn-rpo} are modeled in the SMT encoding.
Similarly, for the value criterion \ref{sn-vc} first all possible
projections are computed. Then an SMT encoding based on the given
rules and theory is constructed and
a model of the encoding (if it exists) delivers
suitable projections that establish termination.
An explicit boolean flag in the SMT encoding determines if a strict or
weak decrease is achieved. The special variant with projections
to suitable linear combinations \ref{sn-vclin}
encodes this by attaching unknown constants to the projected arguments and
summing them up. Those unknowns are then determined by the SMT solver. The
(special) value criterion is currently restricted to the theory of
integers as suitable well-founded orderings are required. For the
integer theory we use $n \succ m$ if $n > m \land n \geqslant 0$ holds.

Method \ref{sn-rpair} receives a DP problem as input and tries to
transform it into a smaller one by orienting strictly as many dependency
pairs as possible. It is parameterized by a list of termination methods
which are applied on the DP problem. The first one
which succeeds determines the remaining problem to be solved. Before
trying to solve the latter, \ref{sn-dpg} is used to decompose it into
smaller problems.

\paragraph{Features.}
Via the command-line arguments several features of \crest{} can be
accessed. This includes control over the number of threads in the
concurrent setup, the overall timeout of the analysis, or if proof output
and debug output should be printed.
Furthermore, (parallel) critical pairs or the dependency graph
approximation of a given LCTRS problem can be computed.
The interface also offers a way to transform an LCTRS into a fully sorted
LCTRS in the ARI format.
In order to alter the default strategy for the analysis, \crest{}
offers a very basic strategy language to specify which methods should be
used. Detailed information is provided in the usage
information of the supplemented artifact.

\section{Improving the Analysis via Transformations}
\label{sec:transformation}
\label{sec:improve-analysis}

In this section we present new transformations which are especially
useful for confluence analysis. These transformations operate on
either rules or constrained critical pairs and
split or unify those based on their constraints.

\subsection*{Splitting Constrained Critical Pairs}
\label{ssec:rule-ccp-split}

If a constrained critical pair has more than one instance, which is almost
always the case, and they cannot all be rewritten by a single rule,
then we are not able to perform any rewrite step. To overcome this problem
we propose a simple method to split constrained critical pairs.

\begin{definition}
\label{def:split-ccp}
Given an \textup{LCTRS} $\xR$, a constrained critical pair
$\rho\colon s \approx t~\CO{\varphi} \in
\CCP(\xR)$ and a constraint $\psi \in \xT(\xFTh,\Var(\varphi))$, the set
$\CCP(\xR)_\rho^\psi$ is defined as
$(\CCP(\xR) \setminus \SET{\rho}) \cup
\SET{s \approx t~\CO{\varphi \land \psi},
s \approx t~\CO{\varphi \land \lnot \psi}}$.
\end{definition}

The following key lemma states that after splitting critical pairs,
all confluence methods are still available. The proof is
given in the appendix.

\begin{lemma}
\label{lem:ccp split}
If $t \Lb[\xR] s \Rb[\xR] u$ then $t \Jb[\xR] u$ or
$t \C_{\CCP_\rho^\psi(\xR)} u$.
\end{lemma}

We illustrate the lemma on the LCTRS in \Cref{exa:example 3}.

\begin{example}
Consider the CCP $\m{g}(x) \approx \m{h}(x)~\CO{\varphi}$
with $\varphi\colon x \geqslant \m{1} \land x \leqslant \m{2}$
from \Cref{exa:example 3}. It is neither in normal form nor trivial.
Since the subterm $\m{g}(x)$ matches the left-hand side of
the rule $\m{g}(x) \R \m{a}~\CO{x = \m{1}}$ (which is how \crest{} renders
the rule $\m{g}(\m{1}) \R \m{a}$), and the combined constraint
$\varphi \land x = \m{1}$ is satisfiable, the CCP is split into
\begin{gather*}
\m{g}(x) \approx \m{h}(x)~\CO{\varphi \land x = \m{1}}
\qquad\text{and}\qquad
\m{g}(x) \approx \m{h}(x)~\CO{\varphi \land x \neq \m{1}}
\end{gather*}
The left one rewrites to the trivial constrained equation
$\m{a} \approx \m{a}~\CO{\varphi \land x = \m{1}}$ using the
rules $\m{g}(x) \R \m{a}~\CO{x = \m{1}}$ and
$\m{h}(x) \R \m{a}~\CO{x \leqslant \m{1}}$.
The right one is rewritten to
$\m{b} \approx \m{b}~\CO{\varphi \land x \neq \m{1}}$ using the rules
$\m{g}(x) \R \m{b}~\CO{x \leqslant \m{2}}$ and
$\m{h}(x) \R \m{b}~\CO{x > \m{1}}$.
Hence the LCTRS $\xR$ is locally confluent by~\Cref{lem:ccp split}.
Using RPO with the precedence
$\m{f} > \m{g} > \m{h} > \m{a} > \m{b} > \m{c}$,
termination of $\xR$ is easily shown and hence $\xR$ is confluent.
\end{example}

The following example shows that constrained critical pairs may be
split infinitely often before local confluence can be verified.

\begin{example}
\label{exa:infinite splitting}
Consider the LCTRS $\xR$ over the theory $\m{Ints}$ consisting of the rules
\begin{alignat*}{4}
\m{a} &\R \m{f}(n) &~& \CO{n \geqslant \m{0}} &
\m{a} &\R \m{g}(n) &~& \CO{n \geqslant \m{0}} \\
\m{f}(n) &\R \m{b} && \CO{n = \m{0}} &
\m{g}(n) &\R \m{b} && \CO{n = \m{0}} \\
\m{f}(n) &\R \m{f}(m) && \CO{n > \m{0} \land \m{2}*m = n} \qquad\qquad &
\m{g}(n) &\R \m{g}(m) && \CO{n > \m{0} \land \m{2}*m = n} \\
\m{f}(n) &\R \m{f}(m) && \CO{n > \m{0} \land \m{2}*m+\m{1} = n} &
\m{g}(n) &\R \m{g}(m) && \CO{n > \m{0} \land \m{2}*m+\m{1} = n}
\end{alignat*}
This LCTRS has a constrained critical pair
$\m{f}(n) \approx \m{g}(m)~\CO{n \geqslant \m{0} \land m \geqslant \m{0}
\land n = n \land m = m}$ originating from $\m{a}$.
To show confluence of $\xR$ we would need to split the pair
in order to make rules applicable for joining a
specific instance. However, there are infinitely many instances with
pairwise different joining sequences.
\end{example}

The next example shows that splitting also helps to
prove non-confluence.

\begin{example}
\label{exa:example 6}
Consider the LCTRS $\xR$ in \Cref{exa:example 3}. By changing the
constraint of the rule $\m{f}(x) \R \m{g}(x)~\CO{x \geqslant \m{1}}$ to
$\CO{x \geqslant \m{0}}$ we obtain a non-confluent LCTRS.
This is shown by splitting the constrained critical pair
$\m{g}(x) \approx \m{h}(x)~\CO{x \geqslant \m{0} \land x \leqslant \m{2}}$,
and subsequently showing that
$\m{g}(x) \approx \m{h}(x)~\CO{x < \m{1} \land x \geqslant \m{0}
\land x \leqslant \m{2}}$ rewrites to the non-trivial normal form
$\m{c} \approx \m{a}~\CO{x < \m{1} \land x \geqslant \m{0} \land
x \leqslant \m{2}}$.
\end{example}

\subsection*{Merging Constrained Rewrite Rules}
\label{ssec:merge-rules}

Next we discuss the merging of constrained rewrite rules. The idea
here is that rewrite steps may become possible after merging similar
rules.

\begin{definition}
\label{def:merge-rules}
Let $\rho_i\colon \ell_i \R r_i~\CO{\varphi_i}$ for $i = 1, 2$ be
variable-disjoint rewrite
rules in an \textup{LCTRS} $\xR$. Suppose
there exists a renaming $\sigma$ such that
$\ell_1 = \ell_2\sigma$, $r_1 = r_2\sigma$ and
$\Var(\varphi_1) = \Var(\varphi_2\sigma)$.
The \textup{LCTRS} $\xR_{\rho_1}^{\rho_2}$ is defined as
\begin{gather*}
(\xR \setminus \SET{\rho_1, \rho_2}) \cup
\SET{\ell_1 \R r_1~\CO{\varphi_1 \lor \varphi_2\sigma}}
\end{gather*}
\end{definition}

The easy proof of the following lemma is omitted.

\begin{lemma}
\label{lem:rule-merge}
The relations $\RbR$ and $\Rb[\xR_{\rho_1}^{\rho_2}]$ coincide.
\end{lemma}

\begin{example}
\label{exa:example 7}
The LCTRS $\xR$ over the theory $\m{Ints}$ consisting of the
rewrite rules
\begin{align*}
\m{f}(x) &\R \ML[7]{$\m{2}$}~\CO{\m{1} \leqslant x \land
x \leqslant \m{3}} &
\m{g}(x) &\R \m{h}(x) &
\m{h}(x) &\R y~\CO{x = \m{2} \land y = x} \\
\m{f}(x) &\R \ML[7]{$\m{g}(x)$}~\CO{\m{2} \leqslant x \land
x \leqslant \m{4}} &&&
\m{h}(x) &\R y~\CO{x = \m{3} \land y = \m{2}}
\end{align*}
admits the constrained critical pair
$\m{2} \approx \m{g}(x)~\CO{\m{1} \leqslant x \land x \leqslant \m{3} \land
\m{2} \leqslant x \land x \leqslant \m{4}}$. After rewriting the subterm
$\m{g}(x)$ to $\m{h}(x)$, no further step is possible because
the rewrite rules for $\m{h}$ are not applicable. However, if we
merge the two rules for $\m{h}$ into
\begin{align*}
\m{h}(x) \R y~\CO{(x = \m{2} \land y = x) \lor (x = \m{3} \land y = \m{2})}
\end{align*}
we can proceed as
$\m{1} \leqslant x \land x \leqslant \m{3} \land \m{2} \leqslant x \land
x \leqslant \m{4}$ implies $((x = \m{2} \land y = x) \lor
(x = \m{3} \land y = \m{2}))\sigma$ for $\sigma(y) = \m{2}$.
This is exactly how \crest{} operates.
\end{example}

\section{Experimental Evaluation}
\label{sec:evaluation}

In this section we show the progress of \crest{} since the start
of its development in early 2023. Initial experiments of an early
prototype of \crest{} were reported in~\cite{SM23}. In the following
tables the prototype of \cite{SM23} is denoted by \crestcade{}. Since then
more criteria for (non-)confluence and termination were added, and
parts of the tool infrastructure were completely revised.
Detailed results are available from the
\href{http://cl-informatik.uibk.ac.at/software/crest}{website} of \crest{}
and the artifact of the experiments at the Zenodo
repository~\cite{crest2024}.

All experiments were performed using the
\texttt{benchexec}\footnote{\url{https://github.com/sosy-lab/benchexec/}}
benchmarking framework
which is also used in the StarExec cluster. The benchmark hardware
consists of several Intel Xeon E5-2650 v4 CPUs having a base clock speed
of 2.20GHz, amounting in total to 64 cores and 128 GB of RAM.
As benchmarks we use the problems in the
new ARI database\footnote{\url{https://ari-cops.uibk.ac.at/ARI/}}
in addition to the examples from this paper.

\paragraph{Tool Setup.}
Each tool receives an ARI benchmark as input and should return
either "YES" (property was proved), "NO" (property was disproved) or
"MAYBE" (don't know) as the first line of its output. In the
tables we represent those with \Y, \N{} and \M{}, respectively.
The fourth category depicted by \T{} denotes that the translation from
the ARI format to the input format of the respective tool failed.
In order to have a realistic setup, a tool has 4 cores including 8 GB of
RAM available for each run. Each tool has 60 seconds to solve a problem
before it is killed. Since we have no information about how many
threads the other tools use, in the experiments we use CPU time over
wall-clock time in order to have a fair comparison.

Our tool \crest{} is split into different binaries depending on the
analysis. The most important ones are \texttt{crest-cr} for confluence and
\texttt{crest-sn} for termination. We use those two including an
additional flag to allow at most 8 threads for the concurrent setup. The
default strategy for confluence uses all methods concurrently and where
specific methods are tested we restrict to those using our strategy flag.
For the default termination setup we use reduction pairs including
dependency graph analysis, recursive path order, (special) value
criterion and subterm criterion.

\Cora, \Ctrl{} and the prototype of~\cite{SM23} do not accept the
ARI format as input. We have developed transformation tools which (try to)
transform an ARI benchmark into their respective input. This might not
always be possible, hence the transformation tool might fail, which is the
reason why we do not distinguish tool (parse) errors from "MAYBE".

\begin{table}[t]
\renewcommand{\arraystretch}{1.25}
\centering
\caption{Confluence analysis of examples.}
\label{tab:examples}
\begin{tabular}{@{}l@{\quad}%
c@{\quad}c@{\quad}c@{\quad}c@{\quad}c@{\quad}c@{}}
\toprule
tool
& \ref{exa:prelim-rewriting}
& \ref{exa:prelim-confluence}
& \ref{exa:example 3}
& \ref{exa:infinite splitting}
& \ref{exa:example 6}
& \ref{exa:example 7} \\ \midrule
\crest  & \Y & \Y & \Y & \M & \N & \Y \\
\CRaris & \M & \M & \M & \M & \M & \M \\
\Ctrl   & \Y & \M & \M & \M & \M & \M \\
\crestcade & \Y & \Y & \M & \M & \M & \M\\
\bottomrule
\end{tabular}
\end{table}

\paragraph{Examples.}
In \Cref{tab:examples} we compare the LCTRS confluence tools 
on the examples in this paper. \crest{} only fails on
\Cref{exa:infinite splitting}, which is a confluent LCTRS, but for which
no automatable method is known.

\paragraph{Confluence Competition.}
The last (2024) Confluence Competition%
\footnote{\url{https://project-coco.uibk.ac.at/2024/index.php}}
hosted the first LCTRS category, with
\crest{} and \CRaris{} as participants.
The former achieved 67 confluence and 26 non-confluence proofs
on a total of 100 selected problems from the ARI
database. \CRaris, which does not (yet) implement techniques for
non-confluence achieved 54 confluence proofs. Currently, \crest{} is the
only tool utilizing a criterion for non-confluence of LCTRSs.

\begin{figure}[b]
\centering
\begin{Verbatim}[fontsize=\small]
(format LCTRS :smtlib 2.6)
(theory Ints)
(fun f (-> Int Int))
(fun g (-> Int Int Int))
(fun a Int)
(rule (f a) (g 4 4))
(rule a (g (+ 1 1) (+ 3 1)))
(rule (g x y) (f (g z y)) :guard (= z (- x 2)))
\end{Verbatim}
\vspace{-2ex}
\caption{ARI file \href{https://ari-cops.uibk.ac.at/ARI/%
?m=problems\&d=ARI\&q=1529}{1529} (without sort annotations and meta
information).}
\label{fig:ari-pcp}
\end{figure}

\paragraph{Confluence.}
All confluence criteria implemented in \crest{}, except \ref{cr-kb},
require left-linearity. For \ref{cr-sc} right-linearity is also required.
Left- and right-linearity is checked only on the non-logical variables.
\Cref{tab:confluence-crest} presents a summary of the confluence methods
implemented in \crest{}. The full set of benchmarks consists of the 107
problems in the ARI database. \crest{} can prove in a full run with all
methods enabled 72 confluent and
26 non-confluent. Of the remaining 9 problems, 2 result in "MAYBE" and 7
in a timeout.
Interesting to observe is that (almost) development
closedness is way faster than (almost) parallel closedness,
which may be due to the fact that less multi-steps than parallel steps
are needed to turn a constrained critical pair into a trivial one.
The number 72 is explained by the fact that \ref{cr-adc} and \ref{cr-pcpcp}
are incomparable: \ref{cr-adc} succeeds on the problem in
\Cref{fig:ari-adc} but fails on the one in \Cref{fig:ari-pcp}, while the
opposite holds for \ref{cr-pcpcp}.

\begin{table}[t]
\renewcommand{\arraystretch}{1.25}
\centering
\caption{Confluence analysis using methods in \crest{} on 107 LCTRSs.}
\label{tab:confluence-crest}
\begin{tabular}{@{}l@{\quad}r@{\quad}r@{\quad}r@{}}
\toprule
criterion & solved & time (AVG) & time (total) \\ \midrule
termination and joinable critical pairs \ref{cr-kb} & 50 & 4.55~s & 487~s
\\
orthogonality \ref{cr-wo} & 62 & 0.10~s & 11~s \\
weak orthogonality \ref{cr-wo} & 65 & 0.12~s & 13~s \\
strongly closed critical pairs \ref{cr-sc} & 56 & 1.21~s & 129~s \\
parallel closed critical pairs \ref{cr-apc} & 66 & 0.44~s & 47~s \\
almost parallel closed critical pairs \ref{cr-apc} & 70 & 11.03~s &
1180~s \\
development closed critical pairs \ref{cr-adc} & 66 & 0.39~s & 42~s \\
almost development closed critical pairs \ref{cr-adc} & 71 & 2.06~s &
220~s \\
parallel closed parallel critical pairs \ref{cr-pcpcp} & 71 & 13.93~s &
1490~s \\
\midrule
all confluence methods \ref{cr-wo}--\ref{cr-pcpcp} & 72 & 8.40~s & 899~s \\
non-confluence (\Cref{lem:non-confluence}) & 26 & 1.96~s & 210~s \\
methods \ref{cr-wo}--\ref{cr-pcpcp} +
(\Cref{lem:non-confluence}) & 98 & 1.84~s & 197~s \\ \midrule
total solved & 98 & --- & --- \\
\bottomrule
\end{tabular}
\end{table}

\begin{table}[b]
\renewcommand{\arraystretch}{1.25}
\centering
\caption{Confluence analysis of LCTRS tools on 107 LCTRSs.}
\label{tab:confluence-tools}
\begin{tabular}{@{}l@{\quad}r@{\quad}r@{\quad}r@{\quad}r@{\quad}r@{\quad}%
r@{\quad}r@{}}
\toprule
tool & \Y & \N & \M & \T & solved & time (AVG) & time (total) \\ \midrule
\CRaris & 58 & 0 & 49 & --- & 54~\% & 0.13~s & 14~s \\
\crest & 72 & 26 & 9 & --- & 92~\% & 1.84~s & 197~s \\
\Ctrl & 54 & 0 & 49 & 4 & 50~\% & 0.17~s & 18~s \\
\crestcade & 67 & 0 & 37 & 3 & 63~\% & 1.14~s & 122~s \\ \midrule
total solved & 72 & 26 & --- & --- & 92~\% & --- & --- \\
\bottomrule
\end{tabular}
\end{table}

In \Cref{tab:confluence-tools} we compare all confluence tools on
the same 107 LCTRS problems. \Ctrl{} supports only weak orthogonality
and \CRaris{} in addition the Knuth--Bendix criterion.
Overall \crest{} is able to solve 92~\% of the LCTRS problems in the
current ARI database and this percentage is reached even if the
timeout is restricted to 10 seconds.
The prototype of~\cite{SM23} supports the methods \ref{cr-wo},
\ref{cr-sc}, \ref{cr-apc} and proves 67 (63~\%) confluent within 122
seconds.

\paragraph{Termination.}
In \Cref{tab:termination-crest} we compare the different
termination methods in \crest. The "dependency graph" method corresponds
to \ref{sn-dpg} with a check for the absence of SCCs,
"recursive path order" corresponds to
\ref{sn-rpo}, "subterm criterion" to \ref{sn-sc}, "(special) value
criterion"
to \ref{sn-vc} (\ref{sn-vclin}) and "reduction pairs" to \ref{sn-rpair}.
The methods annotated with \ref{sn-dpg} work on DP problems and are
applied after an initial dependency graph analysis.
The method "reduction pairs no SVC" uses
\ref{sn-rpo}, \ref{sn-vc} and \ref{sn-sc} and "default" includes
additionally \ref{sn-vclin}. The latter constitutes the current
default setup in \crest.

\begin{table}[tb]
\renewcommand{\arraystretch}{1.25}
\centering
\caption{Termination analysis using methods in \crest{} on 107 LCTRSs.}
\label{tab:termination-crest}
\begin{tabular}{@{}l@{\quad}r@{\quad}r@{\quad}r@{}}
\toprule
method & solved & time (AVG) & time (total) \\ \midrule
DP graph \ref{sn-dpg} & 9 & 0.08~s & 9~s \\
recursive path order \ref{sn-rpo} & 27 & 0.11~s & 12~s \\
recursive path order \ref{sn-dpg}, \ref{sn-rpo} & 28 & 0.11~s & 12~s \\
subterm criterion \ref{sn-dpg}, \ref{sn-sc} & 12 & 0.12~s & 13~s \\
value criterion \ref{sn-dpg}, \ref{sn-vc} & 34 & 0.13~s & 14~s \\
special value criterion \ref{sn-dpg}, \ref{sn-vclin} & 70 & 0.12~s &
13~s \\
reduction pairs no SVC \ref{sn-dpg}--\ref{sn-sc} & 37 & 0.14~s & 15~s \\
default \ref{sn-dpg}--\ref{sn-vclin} & 74 & 0.15~s & 16~s \\ \midrule
total solved & 74 & --- & --- \\
\bottomrule
\end{tabular}
\end{table}

We continue the evaluation by comparing \crest{} to other termination
tools for LCTRSs.
For this comparison we use the higher-order tool \Cora{}
and \Ctrl. The experiments in \Cref{tab:termination} show
that the tools are comparable in strength on the LCTRS benchmarks in
the ARI database,
which is not that surprising as the implemented methods are similar.
All tools together prove 73~\% of the LCTRSs
\begin{table}[b]
\renewcommand{\arraystretch}{1.25}
\centering
\caption{Termination analysis of LCTRS tools on 107 LCTRSs.}
\label{tab:termination}
\begin{tabular}{@{}l@{\quad}r@{\quad}r@{\quad}r@{\quad}r@{\quad}r@{\quad}%
r@{}}
\toprule
tool & \Y & \M & \T & solved & time (AVG) & time (total) \\ \midrule
\Cora & 71 & 30 & 6 & 66~\% & 2.47~s & 264~s \\
\crest & 74 & 33 & --- & 69~\% & 0.15~s & 16~s \\
\Ctrl & 74 & 29 & 4 & 69~\% & 0.96~s & 103~s \\ \midrule
total solved & 78 & --- & --- & 73~\% & --- & --- \\
\bottomrule
\end{tabular}
\end{table}
in the ARI database
terminating. All those tools fail on the bit vector problem
in \Cref{fig:ari-craris} whereas \CRaris{} is able to prove termination
(Naoki Nishida, personal communication).
A fork of the official version of \Ctrl{}%
\footnote{\url{https://github.com/bytekid/ctrl}}
implements the technique of
\cite{NW18} for non-termination of LCTRSs.
Initial experiments reveal that it succeeds
to prove non-termination of 8 problems in~\Cref{tab:termination}.

\begin{figure}[t]
\centering
\begin{Verbatim}[fontsize=\small]
(format LCTRS :smtlib 2.6)
(theory FixedSizeBitVectors)
(fun cnt (-> (_ BitVec 4) (_ BitVec 4)))
(fun u1 (-> (_ BitVec 4) (_ BitVec 4) (_ BitVec 4) (_ BitVec 4)))
(rule (cnt x) (u1 x #b0000 #b0000) )
(rule (u1 x i z) (u1 x (bvadd i #b0001) (bvadd z #b0001))
  :guard (bvult i x)))
(rule (u1 x i z) z :guard (not (bvult i x))))
\end{Verbatim}
\vspace{-2ex}
\caption{ARI file \href{https://ari-cops.uibk.ac.at/ARI/%
?m=problems\&d=ARI\&q=1605}{1605} (without sort annotations and meta
information).}
\label{fig:ari-craris}
\end{figure}

\paragraph{Term Rewrite Systems.}
In the final experiment we compare \crest{} with the state-of-the-art
in automated confluence proving for TRSs. After parsing an input TRS,
\crest{} attaches a single sort to all function symbols and variables,
and adds an empty constraint to all rules. At this point the TRS can be
analyzed as an LCTRS. We compare \crest{} to the latest
winner of the TRS category in the Confluence Competition,
\CSI~\cite{NFM17}, on the 
566 TRS benchmarks in the ARI database.
The results can be seen in \Cref{tab:confluence-trss}.
Keeping in mind that there is some overhead in the analysis of \crest{}
on TRSs as all its methods are geared towards the constrained setting,
the 31\,\% mark is not a bad result. Here it is important to note that
\CSI{} has been
actively developed over a ten-year period and utilizes many more
confluence methods---there is several decades of research on confluence
analysis of TRSs while LCTRS confluence analysis is still in its infancy.

\begin{table}[b]
\renewcommand{\arraystretch}{1.25}
\centering
\caption{Confluence analysis of \crest{} and \CSI{} on 566 TRSs.}
\label{tab:confluence-trss}
\begin{tabular}{@{}l@{\quad}%
r@{\quad}r@{\quad}r@{\quad}r@{\quad}r@{\quad}r@{}}
\toprule
tool & \Y & \N & \M & solved & time (AVG) & time (total) \\ \midrule
\crest & 100 & 73 & 393 & 31~\% & 15.87~s & 8980~s \\
\CSI & 259 & 192 & 115 & 80~\% & 6.25~s & 3540~s \\ \midrule
total solved & 259 & 192 & --- & 80~\% & --- & --- \\
\bottomrule
\end{tabular}
\end{table}

\section{Conclusion and Future Work}
\label{sec:conclusion}

In this paper we presented \crest, an open-source tool for
automatically proving (non-)confluence and termination of LCTRSs. Detailed
experiments were provided to show the power of \crest.

In order to further strengthen the (non-)confluence analysis
in \crest{} we plan to adapt powerful methods like
order-sorted decomposition~\cite{FMZvO15}
and redundant rules~\cite{NFM15,SH24} for plain term rewriting
to the constrained setting. Labeling techniques~\cite{ZFM15} are also
on the agenda. The same holds for termination analysis. Natural
candidates are matrix interpretations~\cite{EWZ2008} as well as
the higher-order methods in \cite{GHKV24}. Especially termination
problems on real values, like the one in \Cref{fig:ari-crest}, should be
supported in future. Also non-termination analysis of
LCTRSs~\cite{NW18} is of interest.
Completion, which is supported in \Ctrl~\cite{WM18}, is another topic for a
future release of \crest{}.
In a recent paper~\cite{ANS24} the semantics of LCTRSs is investigated.
In that context, concepts like checking consistency of constrained
theories are relevant, which are worthy to investigate from an automation
viewpoint.

Since constrained rewriting is highly complex~\cite[Section~3]{SMM24},
a formalization of the implemented techniques in a proof assistant
like Isabelle/HOL is important. The recent advances in the
formalization and subsequent certification of advanced confluence
techniques~\cite{KM23a,KM23b,HKST24} for plain rewriting in connection
with the transformation in \cite[Section 4]{SMM24} make this a realistic
goal.

Finally, to improve the user experience we aim at a convenient web
interface and a richer command-line strategy.

\paragraph{Code and Availability Statement.}
The source code and data that support the contributions of this work are freely
available in the Zenodo repository ``crest - Constrained REwriting Software
Tool: Artifact for TACAS 2025'' at
\url{https://doi.org/10.5281/zenodo.13969852}~\cite{crest2024}.  The authors
confirm that the data supporting the findings of this study are available within
the paper and the artifact.

\begin{figure}[t]
\centering
\begin{Verbatim}[fontsize=\small]
(format LCTRS :smtlib 2.6)
(theory Reals)
(fun sumroot (-> Real Real))
(fun sqrt (-> Real Real))
(rule (sumroot x) 0.0 :guard (>= 0.0 x))
(rule (sumroot x) (+ (sqrt x) (sumroot  (- x 1.0)))
  :guard (not (>= 0.0 x)))
\end{Verbatim}
\vspace{-2ex}
\caption{ARI file \href{https://ari-cops.uibk.ac.at/ARI/%
?m=problems\&d=ARI\&q=1549}{1549} (without meta information).}
\label{fig:ari-crest}
\end{figure}

\begin{credits}
\subsubsection{\ackname}
We thank Fabian Mitterwallner for valuable discussions on automation.
We are grateful to the authors of the tools used in the
experiments for their help in obtaining executables and useful
insights about their usage. The insightful comments and suggestions
provided by the reviewers greatly improved the presentation of the paper.

\subsubsection{\discintname}
The authors have no competing interests to declare that are
relevant to the content of this article.
\end{credits}

\bibliographystyle{splncs04}
\bibliography{references}

\appendix

\section{Appendix}

In this appendix we state the closing conditions on (parallel)
critical pairs that are used in the confluence results implemented in
\crest{} and we present the proofs of Lemmata~\ref{lem:non-confluence}
and~\ref{lem:ccp split}.

\begin{definition}
\label{def:sc}
A constrained critical pair $s \approx t~\CO{\varphi}$ is
\emph{strongly closed} if
\begin{enumerate}
\item
$s \approx t~\CO{\varphi}
\mr{\,\Rs_{\geqslant 1}^* \cdot \,\Rs_{\geqslant 2}^=\,}
u \approx v~\CO{\psi}$ for some trivial $u \approx v~\CO{\psi}$, and
\item
$s \approx t~\CO{\varphi}
\mr{\,\Rs_{\geqslant 2}^* \cdot \,\Rs_{\geqslant 1}^=\,}
u \approx v~\CO{\psi}$ for some trivial $u \approx v~\CO{\psi}$.
\end{enumerate}
An LCTRS is strongly closed if all its constrained critical pairs are
strongly closed.
\end{definition}

\begin{definition}
\label{def:dc}
A constrained critical pair $s \approx t~\CO{\varphi}$ is
\emph{development closed} if
$s \approx t~\CO{\varphi} \sMSab[\geqslant 1]{} u \approx v~\CO{\psi}$
for some trivial $u \approx v~\CO{\psi}$.
A constrained critical pair is \emph{almost development closed} if it is
not an overlay
and development closed, or it is an overlay and
$s \approx t~\CO{\varphi} \sMSab[\geqslant 1]{} \cdot \sRab[\geqslant 2]{*}
u \approx v~\CO{\psi}$ for some trivial $u \approx v~\CO{\psi}$.
An LCTRS is called (almost) development closed if all its
constrained critical pairs are (almost) development closed.
\end{definition}

In the following we denote for a term $s$, a set of parallel
positions $P$ in $s$, and a set of terms
$\SET{t_p}_{p \in P}$ by $s[t_p]_{p \in P}$
the simultaneous replacement of $s|_p$ in $s$ by $t_p$ for all
$p \in P$.  The notion $\TVar(s,\varphi,P)$ in the variable condition of
the next definition expands to
$\bigcup_{p \in P} \Var(s|_p) \setminus \Var(\varphi)$.

\begin{definition}
\label{def:pc}
A constrained critical pair $s \approx t~\CO{\varphi}$ is
\emph{1-parallel closed} if $s \approx t~\CO{\varphi}
\sptoab[\geqslant 1]{} \cdot \sRab[\geqslant 2]{*} u \approx v~\CO{\psi}$
for some trivial $u \approx v~\CO{\psi}$. An LCTRS is 1-parallel
closed if all its constrained critical pairs are 1-parallel closed. A
constrained parallel critical pair
$\ell\sigma[r_p\sigma]_{p \in P} \approx r\sigma~\CO{\varphi}$
is \emph{2-parallel closed} if there exists a set of parallel positions
$Q$ such that
\begin{align*}
\ell\sigma[r_p\sigma]_{p \in P} \approx r\sigma~\CO{\varphi}
\sptoab[\geqslant 2]{Q} \cdot \sRab[\geqslant 1]{*} u \approx v~\CO{\psi}
\end{align*}
for some trivial $u \approx v~\CO{\psi}$ and
$\TVar(v,\psi,Q) \subseteq \TVar(\ell\sigma,\varphi,P)$.
An LCTRS is 2-parallel closed if all its constrained parallel
critical pairs are 2-parallel closed. An LCTRS is parallel closed
if it is 1-parallel closed and 2-parallel closed.
\end{definition}

\begin{proof}[of \Cref{lem:non-confluence}]
Assume an LCTRS $\xR$ and a constrained critical pair
$s \approx t~\CO{\varphi} \in \CCP(\xR)$. Further assume that it
rewrites to the non-trivial normal form $u \approx v~\CO{\varphi}$. There
exists a substitution $\sigma \vDash \varphi$ such that
$t\sigma \neq u\sigma$ and
$t\sigma \approx u\sigma$ is a normal form. By definition
of constrained critical pair there exists a term $s$ with
$t\sigma \La[*] s \Ra[*] u\sigma$, which shows non-confluence of $\xR$.
\end{proof}

\begin{proof}[of \Cref{lem:ccp split}]
Assume an LCTRS $\xR$ and let $t \Lb[\xR] s \Rb[\xR] u$.
By the critical pair lemma~\cite[Lemma~20]{WM18} we have either
$t \Jb[\xR] u$ or $t \C_{\CCP(\xR)} u$. It remains to show that for terms
$s_1, s_2\in \xT(\xF,\xV)$ if
$s_1 \C_{s \,\approx\, t\,\CO{\varphi}} s_2$ then either
$s_1 \C_{s \,\approx\, t\,\CO{\varphi \land \psi}} s_2$ or
$s_1 \C_{s \,\approx\, t\,\CO{\varphi \land \lnot \psi}} s_2$.
So suppose
$s_1 \C_{s \,\approx\, t\,\CO{\varphi}} s_2$.
There exist a context $C$
and a substitution $\sigma$ such that
$s_1 = C[s\sigma]$, $s_2 = C[t\sigma]$ and $\sigma \vDash \varphi$.
From $\Var(\psi) \subseteq \Var(\varphi)$ we obtain
$\sigma \vDash \varphi \land \psi$ or
$\sigma \vDash \varphi \land \lnot \psi$.
Hence $t \Jb[\xR] u$ or $t \C_{\CCP_\rho^\psi(\xR)} u$.
\end{proof}

\end{document}